\documentclass[preprintnumbers,amsmath,amssymb]{revtex4}
\def\version{February 25, 2010}

\usepackage{upgreek}
\usepackage{latexsym,epsfig,bm,amssymb}
\usepackage{color}
\usepackage{amsthm,mathrsfs}
\usepackage{amsmath}
\usepackage{mathrsfs}
\usepackage{times}

\nonstopmode

\newcommand{\notyet}[1]{}

\DeclareSymbolFont{AMSb}{U}{msb}{m}{n}
\DeclareSymbolFontAlphabet{\mathbb}{AMSb}

\newcommand{\h}{{h^{\hspace{-2.5mm}-}}}

\newcommand{\X}{\mathscr{X}}

\newcommand{\Y}{\mathscr{Y}}
\newcommand{\bS}{\mathbf{S}}

\newcommand{\dist}{\,{\rm dist}\,}

\newcommand{\supp}{\mathop{\rm supp}}

\newcommand{\p}{\partial}
\newcommand{\at}[1]{\vert\sb{\sb{#1}}}

\def\Re{{\rm Re\, }}
\def\Im{{\rm Im\,}}
\providecommand{\C}{\mathbb{C}}
\renewcommand{\C}{\mathbb{C}}
\newcommand{\R}{\mathbb{R}}
\newcommand{\N}{\mathbb{N}}

\newcommand{\abs}[1]{\vert #1 \vert}

\newcommand{\norm}[1]{\Vert #1 \Vert}

\newcommand{\sothat}{{\rm :}\ }

\providecommand{\ltor}[1]{
\ifnum #1=1{\it i}\else\ifnum #1=2{\it ii}\else\ifnum #1=3{\it iii}
\else\ifnum #1=4 {\it iv}\fi\fi\fi\fi
}

\DeclareMathSymbol{\varGamma}{\mathord}{letters}{"00}
\DeclareMathSymbol{\varDelta}{\mathord}{letters}{"01}
\DeclareMathSymbol{\varSigma}{\mathord}{letters}{"06}
\DeclareMathSymbol{\varPhi}{\mathord}{letters}{"08}
\DeclareMathSymbol{\varOmega}{\mathord}{letters}{"0A}

\theoremstyle{plain}
\newtheorem{theorem}{Theorem}[section]

\newtheorem{proposition}[theorem]{Proposition}
\theoremstyle{definition}
\newtheorem{definition}[theorem]{Definition}
\newtheorem{assumption}[theorem]{Assumption}

\newtheorem{remark}[theorem]{Remark}

\theoremstyle{remark}

\makeatletter\@addtoreset{equation}{section}
\makeatother

\begin{document}
\title{On global attraction to quantum stationary states.
\\
Dirac equation with mean field interaction}

\author{Alexander Komech}

\affiliation{
Faculty of Mathematics, University of Vienna, Wien A-1090, Austria}
\altaffiliation[Also at]{
Institute for Information Transmission Problems,
Moscow 101447, Russia}

\author{Andrew Komech}

\affiliation{
Mathematics Department, Texas A\&M University,
College Station, TX, USA}

\altaffiliation[Also at]{
Institute for Information Transmission Problems,
Moscow 101447, Russia}

\date{\version}

\begin{abstract}
We consider a $\mathbf{U}(1)$-invariant nonlinear Dirac equation,
interacting with itself via the mean field mechanism.
We analyze the long-time asymptotics of solutions
and prove that, under certain generic assumptions,
each finite charge solution converges as $t\to\pm\infty$
to the two-dimensional set of all ``nonlinear eigenfunctions''
of the form $\phi(x)e\sp{-i\omega t}$.
This global attraction is caused by the nonlinear energy transfer
from lower harmonics to the continuous spectrum
and subsequent dispersive radiation.

The research
is inspired by Bohr's postulate on
quantum transitions and
Schr\"o\-din\-ger's identification
of the quantum stationary states to the nonlinear eigenfunctions
of the coupled $\mathbf{U}(1)$-invariant
Maxwell-Schr\"o\-din\-ger and Maxwell-Dirac equations.
\end{abstract}

\maketitle


\section{Introduction.
Bohr's transitions as global attraction}

In the present paper we continue our research
on the global attraction to solitary wave solutions in
$\mathbf{U}(1)$-invariant dispersive nonlinear systems.
Our aim is a dynamical interpretation of ``quantum jumps''
postulated by N. Bohr in 1913 to explain the stability
of the Rutherford model of the atom (1911).

The long time asymptotics for
nonlinear field theory equations,
and in particular the nonlinear Klein-Gordon equation,
have been the subject of intensive research,
starting with the pioneering papers by
Segal
\cite{MR0153967},
Strauss \cite{MR0233062},
and Morawetz and Strauss \cite{MR0303097},
where the nonlinear scattering
and the local attraction to zero solution were considered.
The orbital stability
of solitary wave solutions of nonlinear Schr\"odinger
and Klein-Gordon equations
is well-understood (see \cite{MR901236}),
and there is presently
an active research
on local attraction to solitary waves,
or \emph{asymptotic stability}
\cite{MR1972870,MR2422523}.

Traditionally, the Dirac equation has presented difficulties.
The existence of solitary waves
in the nonlinear Dirac equation
was
proved in \cite{MR847126}.
The existence of solitary waves in the Maxwell-Dirac system
is proved in
\cite{MR1386737}.
The stability of solitary waves with respect
to a particular class of perturbations
was analyzed in \cite{MR848095}.
The spectral stability of small amplitude solitary waves
of a particular nonlinear Dirac equation
in one dimension,
known as the massive Gross-Neveu model
\cite{PhysRevD.10.3235,PhysRevD.12.3880},
has been confirmed numerically in
\cite{MR2217129}.
Neither the orbital stability
nor asymptotic stability
of solitary waves
is presently understood
in the context of the Dirac equation.

The next question
is related to the structure of a global attractor
of all finite energy solutions.
Global attraction to \emph{static},
stationary solutions in dispersive systems
\emph{without $\mathbf{U}(1)$ symmetry}
was first established in
\cite{MR1359949,MR1434147,MR1726676,MR1748357}.
The attraction
of any finite energy solution
to the set of all solitary waves
in the context of the Klein-Gordon equation
coupled to one and to several nonlinear oscillators
was proved in
\cite{ubk-arma,ukk-jmpa}.
In \cite{ukk-jmpa}, we generalized this result
for the Klein-Gordon field coupled to several oscillators.
In \cite{ukr-hp}, a similar result
was generalized to a higher dimensional setting
for the Klein-Gordon equation with the mean field interaction.
In the present paper,
we consider the structure of the attractor
for the Dirac equation,
in the model with the mean field interaction.
We will show that under rather general assumptions
the attractor coincides with the set of all solitary waves.

According to Bohr's postulates \cite{bohr1913},
an unperturbed electron runs forever
along
certain \emph{stationary orbit},
which we denote
$\vert E\rangle$
and call \emph{quantum stationary state}.
Once in such a state,
the electron has a fixed value of energy $E$,
with the energy not being lost via emitted radiation.
Under a perturbation,
the electron can jump from one
quantum stationary state to another,
$
\vert E\sb{-}\rangle
\longmapsto
\vert E\sb{+}\rangle,
$
radiating (or absorbing) a quantum of light.
Bohr's postulate
suggests the dynamical interpretation
of Bohr's transitions
as a long-time asymptotics
\begin{equation}\label{ga}
\psi(t)\longrightarrow\vert E\sb\pm\rangle,
\qquad
t\to\pm\infty
\end{equation}
for any trajectory $\psi(t)$ of the corresponding dynamical system,
where the limiting states $\vert E\sb\pm\rangle$
depend on the trajectory.
Then the
\emph{quantum stationary states}
should be viewed as points of the \emph{global attractor}.


At first glance, the global attraction (\ref{ga})
seems incompatible with the energy conservation and
time reversibility of Hamiltonian systems.
We intend to verify that
such asymptotics in principle are possible
for 
nonlinear Hamiltonian field equations.
In this paper, we verify this asymptotics
for equations of Dirac type.


Developing de Broglie's ideas,
Schr\"odinger
identified quantum stationary states
of energy $E$ with the solutions of type
\begin{equation}\label{qss}
\psi(x,t)=e^{-iEt/\h}\psi(x),
\qquad
x\in\R^3.
\end{equation}
Then the Schr\"odinger equation
\begin{equation}\label{S}
i\hbar\p\sb t\psi
=H\psi:=-\frac{\hbar^2}{2m}\Delta\psi+V(x)\psi,
\qquad
\psi=\psi(x,t)\in\C,
\quad
x\in\R^3,
\end{equation}
becomes the \emph{eigenvalue problem}
$
E\psi(x)=H\psi(x).
$
One of the original  Schr\"odinger's ideas
\cite{sch79361}
was to identify the integers in the Debye-Sommerfeld-Wilson
quantum rules
with the integers arising in the eigenvalue problems for PDEs.

For the case of the {\it free particles},
this identification agrees with the de Broglies
wave function
$\psi(x,t)=e^{i p\cdot x/\h}e^{-iE t/\h}$,
where $p\in\R^3$ is the momentum of the particle.
For the {\it bound particles} in an external potential,
the identification (\ref{qss}) reflects the fact that
the space is ``twisted'' by the external field, while
the time remains ``free''.

Thus, the reason for Schr\"odinger's choice
of quantum stationary states
in the form (\ref{qss})
seems to be rather
algebraic.
At the same time,
the attraction \eqref{ga}
suggests the dynamical interpretation
of the quantum stationary states
as asymptotic states in the long time limit,
that is,
\begin{equation}\label{asymptotics}
\psi(x,t)
\sim
\phi\sb{\omega\sb\pm}(x)e\sp{-i\omega\sb{\pm}t},
\qquad
t\to\pm\infty.
\end{equation}
Such asymptotics should hold for each finite charge solution.
The asymptotics of type \eqref{asymptotics}
are generally impossible for
the linear autonomous
equation (\ref{S}) because of the superposition
principle.
The asymptotics would
mean that
the solitary waves
(\ref{qss}) form the global attractor for the
coupled nonlinear
Maxwell-Schr\"odinger and Maxwell-Dirac systems.

An adequate description of
``quantum jumps''
in an atom
requires that we consider the equation
for the electron wave function
(Schr\"o\-din\-ger or Dirac equation)
coupled to the Maxwell system
which governs the time evolution of the four-potential
$A\sp\mu(x,t)=(\varphi(x,t),\mathbf{A}(x,t))$.
The corresponding
Maxwell-Schr\"odinger system
was initially introduced
by Schr\"odinger
in \cite{Sch81109}.
Its global well-posedness
was considered in \cite{MR1331696}.
The results on local existence of solutions
to the Dirac-Maxwell system
were obtained in \cite{MR1391520}.
The coupled systems are
$\mathbf{U}(1)$-invariant with respect
the (global) gauge group
$\big(\psi(x), A\sp\mu(x)\big)
\mapsto
\big(
\phi(x)e\sp{-i\theta},
A\sp\mu(x)
\big)$, $\theta\in\R$. Respectively,
one might expect the following natural
generalization of asymptotics \eqref{asymptotics}
for solutions to the coupled
Maxwell-Schr\"o\-din\-ger
or Maxwell-Dirac
equations:
\begin{equation}\label{asymptotics-a}
\big(\psi(x,t), A\sp\mu(x,t)\big)
\sim
\big(
\phi\sb\pm(x)e\sp{-i\omega\sb\pm t},
A\sp\mu\sb\pm(x)
\big),
\qquad
t\to\pm\infty.
\end{equation}
The asymptotics \eqref{asymptotics-a}
would mean that
the set of all solitary waves
forms a global attractor for the coupled system.
The asymptotics of this type
are not available yet in the context of coupled systems.


\section{Model and the result}

In the present paper
we consider
the Dirac equation with the mean field interaction:
\begin{equation}\label{nd-mf}
i\dot\psi(x,t)
=\big(-i\sum\sb{j=1}\sp{3}\bm\alpha\sb j\p\sb j+\bm\beta m\big)\psi
+\rho(x)F(\langle\rho,\psi(\cdot,t)\rangle),
\quad
\psi\in\C^4,
\quad
x\in\R^3,
\quad n\ge 1,
\end{equation}
where the Hermitian $4\times 4$ matrices
$\bm\alpha\sb j$ and $\bm\beta$
satisfy
$\{\bm\alpha\sb j,\bm\alpha\sb k\}=2\delta\sb{jk}\bm I$,
$\{\bm\alpha\sb j,\bm\beta\}=\bm 0$,
$\bm\beta^2=\bm I$.
We write
$
\langle\rho,\psi(\cdot,t)\rangle
=\displaystyle\int\sb{\R^3}\rho\sp\dagger(x)\psi(x,t)\,d^3 x,
$
where $\rho$ is a spinor-valued
coupling
function
from the Schwartz class:
\[
\rho
=
{\tiny
\left[\begin{array}{c}\rho\sb 1\\ \dots \\ \rho\sb 4\end{array}
\right]
}
\in\mathscr{S}(\R^3,\C^4),
\qquad
\rho\not\equiv\bm 0,
\qquad
\rho\sp\dagger=[\bar\rho\sb 1,\dots,\bar\rho\sb 4].
\]


\begin{assumption}\label{nd-ass-f}
$F(z)=-\nabla\sb{\Re z,\Im z} U(z)$
with
$U(z)=\sum\limits\sb{k=1}\sp{p}u\sb k\abs{z}\sp{2k}$,
where
$u\sb k\in\R$,
$p\ge 2$,
and
$u\sb p>0$.
\end{assumption}
Since one has
$F(e\sp{i\theta}z)=e\sp{i\theta} F(z)$
for $\theta\in\R$, $z\in\C$,
equation \eqref{nd-mf}
is $\mathbf{U}(1)$-invariant,
where $\mathbf{U}(1)$ stands for the unitary group
$e\sp{i\theta}$, $\theta\in\R\mod 2\pi$.
By the N\"other theorem, the charge functional
$
\mathcal{Q}(\psi)
=\int\sb{\R^3}
\abs{\psi}^2\,d^3 x
$
is conserved.

Equation \eqref{nd-mf} is the simplest model
sharing the following features
with the Maxwell-Dirac system,
which we consider the key for global attraction
to solitary waves:
The model has $\mathbf{U}(1)$-symmetry;
It is dispersive;
It is nonlinear.
The first feature allows for solitary waves,
while
the last two features
are responsible for the ``friction by dispersion''
mechanism:
The nonlinearity moves the perturbations
into the continuous spectrum of the linearized equation,
and thereafter the perturbations are dispersed
to infinity.
Let us also mention
that the modelling of the
interaction of the matter with
gauge fields
in terms of local self-interaction
takes its origin back to
at least as early as
\cite{PhysRev.84.1},
where the Lorentz-invariant
nonlinear Klein-Gordon equation originally appeared.





\begin{definition}
The space of states
of finite charge is
$\X=L\sp 2(\R^3,\C^4)$,
with the standard norm
denoted by
$\norm{\cdot}\sb{L\sp 2}$.
\end{definition}

Equation \eqref{nd-mf}
formally can be written as a Hamiltonian system
with the phase space $\X$,
$
\dot\psi(t)=J\,D\mathcal{H}(\psi),
$
where $D\mathcal{H}$ is the variational
derivative
with respect to $\Re\psi\sb\mu$, $\Im\psi\sb\mu$,
$\mu=1,\,\dots,\,4$,
of the Hamiltonian
\[
\mathcal{H}(\psi)
=\frac 1 2
\int\sb{\R^3}
\psi\sp\dagger
\big(
-i
\bm\alpha\sb j\p\sb j+\bm\beta m
\big)\psi
\,d^3 x
+U(\langle\rho,\psi\rangle),
\qquad
\psi\sp\dagger=[\bar\psi\sb 1,\,\dots,\,\bar\psi\sb 4].
\]

Let
$
\mathscr{D}(\xi)
=\bm\alpha\sb j\xi\sb j+\bm\beta m,
$
$
\xi\in\R^3.
$
Since $\mathscr{D}(\xi)$ is self-adjoint
and $\mathscr{D}(\xi)^2=\xi^2+m^2$,
its eigenvalues are $\pm\sqrt{\xi^2+m^2}$.
Define the projectors
$
\Pi\sb\pm(\xi)
=\frac{1}{2}\Big(1\pm\frac{\mathscr{D}(\xi)}{\sqrt{\xi^2+m^2}}\Big)
$
onto the corresponding eigenspaces,
and denote
$\hat\rho\sb{\pm}(\xi):=\Pi\sb\pm(\xi)\hat\rho(\xi)$.


\begin{assumption}\label{nd-ass-rho}
\begin{enumerate}
\item
For any $\lambda>0$,
$\hat\rho\sb{+}(\xi)$ and
$\hat\rho\sb{-}(\xi)$
do not vanish identically
on the sphere $\abs{\xi}=\lambda$.
\item
The function
\begin{equation}\label{nd-def-sigma}
\sigma(\omega)
=\int\sb{\R^3}
\frac{
\hat\rho\sp\dagger(\xi)
(\omega+\bm\alpha\sb j\xi\sb j+\bm\beta m)\hat\rho(\xi)}
{\omega^2-\xi^2-m^2}
\,\frac{d^3\xi}{(2\pi)^3}
\end{equation}
is nonzero
for all
$\omega\in[-m,m]
$,
except perhaps at one point
which will be denoted
$\omega\sb\sigma\in[-m,m]
$.
\end{enumerate}
\end{assumption}

\begin{remark}
Note
that limits $\sigma(m-0)$, $\sigma(-m+0)$
exist.
\end{remark}

Let us give examples
of such functions $\rho\in\mathscr{S}(\R^3,\C^4)$.
We take a spherically symmetric function $\rho$
such that $\hat\rho(\xi)\ne 0$ for $\xi\ne 0$,
with
$\hat\rho(\xi)$ being an eigenvector of $\bm\beta$
(since $\bm\beta$ is Hermitian and $\bm\beta^2=1$, its eigenvalues are $\pm 1$).
Then
$\abs{\hat\rho\sb\pm(\xi)}^2
=\frac 1 2\hat\rho\sp{\dagger}(\xi)
\big(1\pm\frac{\bm\alpha\sb j\xi\sb j+\bm\beta m}{\sqrt{\xi^2+m^2}}\big)
\hat\rho(\xi)$,
hence
\[
\int\sb{\abs{\xi}=\lambda}\abs{\hat\rho\sb\pm(\xi)}^2\,d^2\Omega
\ge
\int\sb{\abs{\xi}=\lambda}\hat\rho(\xi)\sp{\dagger}
\big(1-\frac{m}{\sqrt{\xi^2+m^2}}\big)
\hat\rho(\xi)\,d^2\Omega>0,
\qquad
\lambda>0.
\]
The integral of the term with $\bm\alpha\sb j\xi\sb j$
dropped out since $\hat\rho(\xi)$ is spherically symmetric.

Let us now consider $\sigma(\omega)$.
Due to $\hat\rho$ being spherically symmetric,
$\bm\alpha\sb j\xi\sb j$-term cancels out
from the integration in \eqref{nd-def-sigma}.
Since we require that
$\hat\rho(\xi)$ be an eigenvector of $\bm\beta$,
one has
$
\hat\rho(\xi)\sp\ast
(\omega+\bm\beta m)\hat\rho(\xi)
=(\omega\pm m)\hat\rho(\xi)\sp\ast\hat\rho(\xi),
$
hence the expression under the integral
in \eqref{nd-def-sigma} is sign-definite
for all $\omega\in[-m,m]$ except at $\omega=m$ or at $\omega=-m$.
Therefore,
$\sigma(\omega)$ does not vanish on $[-m,m]$
except at one of the endpoints.


\begin{definition}
\label{nd-def-solitary-waves}
The solitary waves of equation \eqref{nd-mf}
are solutions of the form
$\psi(x,t)=\phi\sb\omega(x) e\sp{-i\omega t}$,
where
$\omega\in\R$,
$\phi\sb\omega(x)\in\X$.
The solitary manifold is the set
$
\bS
=
\left\{
\phi\sb\omega\sothat
\mbox{$\phi\sb\omega(x)e^{-i\omega t}$
is a solitary wave}
\right\}\subset\X.
$
\end{definition}

Denote $\C\sp{+}=\{\omega\in\C\sothat \Im\omega>0\}$.
For
$\omega\in\C\sp{+}\cup(-m,m)$,
introduce
\begin{equation}\label{nd-def-v-hat}
\hat\varSigma(\xi,\omega)
=
\frac{(\omega+\bm\alpha\sb j\xi\sb j+\bm\beta m)\hat\rho(\xi)}{\omega^2-\xi^2-m^2}
,
\qquad
\varSigma(x,\omega)
=\mathcal{F}^{-1}\sb{\xi\to x}
\big[
\hat\varSigma(\cdot,\omega)
\big].
\end{equation}
$\varSigma(\cdot,\omega)$ is an analytic function of $\omega\in\C\sp{+}$
with the values in $\mathscr{S}(\R^3,\C^4)$.
One can show that
for any $x\in\R^3$,
$\varSigma(x,\omega)$
can be extended
to the real line $\omega\in\R$
as the boundary trace (in the sense of tempered distributions):
\begin{equation}\label{nd-def-v-real}
\varSigma(x,\omega)=\lim\sb{\epsilon\to 0+}\varSigma(x,\omega+i\epsilon),
\qquad\omega\in\R.
\end{equation}
One has
$\varSigma(\cdot,\omega)\in\X$
for $\omega\in(-m,m)$.
For $\omega=\pm m$,
$\varSigma(\cdot,\omega)\in\X$
only when
$\hat\rho\sb\pm(0)=0$.

\begin{proposition}
\label{nd-prop-solitons}
Let Assumptions~\ref{nd-ass-f}
and~\ref{nd-ass-rho} be satisfied.
Then there are no nonzero solitary wave solutions
to \eqref{nd-mf}
for $\omega\notin[-m,m]$.
For $\omega\in(-m,m)$,
there is a nonzero solitary wave
$\phi\sb\omega(x)e^{-i\omega t}$ if and only if
there is $C\in\C\backslash 0$
such that
\begin{equation}\label{nd-cond-0}
-\sigma(\omega)U'(C\sigma(\omega))=1,
\end{equation}
with $U(\cdot)$ from Assumption~\ref{nd-ass-f}.
One has $\phi\sb\omega(x)=C\varSigma(x,\omega)$.
For $\omega=\pm m$,
$\phi\sb\omega(x)\in\X$
only when $\hat\rho\sb{\pm}(0)=0$.
\end{proposition}

\begin{remark}
Due to Assumption~\ref{nd-ass-f},
the set $\bS$
from Definition~\ref{nd-def-solitary-waves}
is invariant under multiplication by $e\sp{i\theta}$,
$\theta\in\R$.
Generically,
the solitary manifold
$\bS$
is two-dimensional.
It may contain disconnected components.
\end{remark}

\begin{remark}
Since $F(0)=0$,
the zero solitary wave
is an element of $\bS$,
formally corresponding to any $\omega\in\R$.
\end{remark}

\begin{remark}
It is possible that
for some $\omega$
equation \eqref{nd-cond-0}
has nonzero roots $C\sb j\in\C$
of different magnitude;
then there are solitary wave solutions
corresponding to each of these roots.
\end{remark}


\begin{definition}
Fix $\varepsilon>0$ and
$\chi\in C\sp\infty\sb{comp}(\R^3)$,
$0\le\chi\le 1$,
$\chi(x)\equiv 1$ for $\abs{x}<1$,
with $\supp\chi$
in the ball of radius $2$.
Denote by $\Y$ the space $\X$
endowed with the metric
$
\norm{\psi}\sb{\Y}
:=\sum\sb{R\in\N}
2^{-R}\norm{\chi(x/R)\psi(x)}\sb{{H\sp{-\varepsilon}}}
$,
where the norm in the Sobolev space $H\sp s$
is given by
$\norm{\psi}\sb{H\sp s}
=\norm{(m^2-\Delta)^{s/2}\psi}\sb{L\sp 2}$.
\end{definition}

The Sobolev Embedding Theorem implies that
the embedding $\X\subset\Y$
is compact
for any $\varepsilon>0$.

\begin{theorem}[Main Theorem]
\label{nd-main-theorem}
Assume that
the nonlinearity $F(z)$
is as in Assumption~\ref{nd-ass-f}
and that
the coupling function $\rho(x)$ satisfies Assumption~\ref{nd-ass-rho}.
Then for any $\psi\sb 0\in \X$
there is a global solution $\psi(t)\in C(\R,\X)$
to equation \eqref{nd-mf},
such that $\psi\at{t=0}=\psi\sb 0$,
which converges to $\bS$ in the space
$\Y$:
\[
\lim\sb{t\to\pm\infty}
\dist\sb{\Y}(\psi(t),\bS)=0,
\qquad
\mbox{where}
\quad
\dist\sb{\Y}(\psi,\bS)
:=\inf\limits\sb{\bm s\in\bS}
\norm{\psi-\bm s}\sb{\Y}.
\]
\end{theorem}


\begin{figure}[htbp]
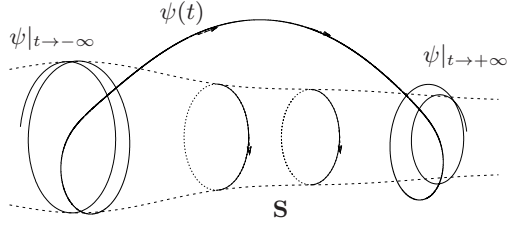

\input fig-attractor-sm.tex
\caption{
For $t\to\pm\infty$,
a finite charge solution $\psi(t)$
approaches the global attractor
which coincides with the set of all solitary waves
$\bS$.}
\label{fig-attractor}
\end{figure}


We work in the $L\sp 2$ setting
because of the absence of existence results
for finite energy solutions.
As a result, the convergence
to the attractor
holds just below $L\sp 2$,
in local $H\sp{-\varepsilon}$-norm.
The convergence to $\bS$
is due to the ``friction by dispersion'':
the nonlinearity carries the excess energy
into the continuous spectrum,
where it is subsequently dispersed to infinity,
and the remaining part of the solution
settles to one of the nonlinear eigenstates
(or zero).
By the Titchmarsh Convolution Theorem
(see \cite{titchmarsh} and \cite{MR617913}),
this process can only stop
when the time-spectrum of the solution
consists of the single frequency,
that is, when the limiting solution
coincides with a nonlinear solitary wave
(a nonlinear Schr\"odinger eigenstate).
It is only for such states
that
the energy transfer
from lower harmonics into the continuous spectrum
(the \emph{spectral inflation})
is absent.

Let us comment on our methods.
We follow the path
developed in \cite{ubk-arma,ukr-hp,ukk-jmpa}:
we prove the absolute continuity of the
spectral density for large frequencies,
use
the compactness argument to extract the omega-limit trajectories,
and show that these trajectories
have the finite time spectrum.
Then we use the Titchmarsh Convolution Theorem
to pinpoint the spectrum to a single frequency.
Let us note that the Titchmarsh theorem
is only applicable
if we assume that $F$
admits a real-valued polynomial potential
(Assumption~\ref{nd-ass-f}).

\section{Conclusion}

We demonstrated that in a particular nonlinear dispersive
equation,
whose linearization is the Dirac equation,
the convergence to Schr\"odinger states
for long positive and negative times
takes place as a consequence of
``friction by dispersion'',
so that the solution
settles to one of the nonlinear eigenstates
(or to zero).
We conjecture that a similar consideration
will allow to prove convergence to nonlinear
eigenstates
in systems such as coupled Maxwell-Dirac equations,
providing a dynamical description
of Bohr's ``quantum jumps''.

\begin{acknowledgments}
We would like to express our gratitude
to
Dmitry Kazakov,
Valery Pokrovsky,
Alexander Shnirelman,
Herbert Spohn, Walter Strauss
for stimulating discussions.
The first author was supported
by Alexander von
Humboldt Research Award (2006), and by grants FWF  P19138-N13,
DFG
436\,RUS\,113/929/0-1, and RFBR.
The second author was supported
by the National Science Foundation under Grant DMS-0600863.
\end{acknowledgments}

\hbadness=10000


\def\cprime{$'$} \def\cprime{$'$} \def\cprime{$'$} \def\cprime{$'$}
  \def\cprime{$'$} \def\cprime{$'$} \def\cprime{$'$} \def\cprime{$'$}
  \def\cprime{$'$} \def\cprime{$'$}
\begin{thebibliography}{Sch26b}

\bibitem[Boh13]{bohr1913}
N.~Bohr, {\em On the constitution of atoms and molecules\/}, Phil. Mag. {\bf
  26} (1913), pp. 1--25.

\bibitem[Bou96]{MR1391520}
N.~Bournaveas, {\em Local existence for the {M}axwell-{D}irac equations in
  three space dimensions\/}, Comm. Partial Differential Equations {\bf 21}
  (1996), pp. 693--720.

\bibitem[BS03]{MR1972870}
V.~S. Buslaev and C.~Sulem, {\em On asymptotic stability of solitary waves for
  nonlinear {S}chr\"odinger equations\/}, Ann. Inst. H. Poincar\'e Anal. Non
  Lin\'eaire {\bf 20} (2003), pp. 419--475.

\bibitem[CP06]{MR2217129}
M.~Chugunova and D.~Pelinovsky, {\em Block-diagonalization of the symmetric
  first-order coupled-mode system\/}, SIAM J. Appl. Dyn. Syst. {\bf 5} (2006),
  pp. 66--83.

\bibitem[Cuc08]{MR2422523}
S.~Cuccagna, {\em On asymptotic stability in energy space of ground states of
  {NLS} in 1{D}\/}, J. Differential Equations {\bf 245} (2008), pp. 653--691.

\bibitem[CV86]{MR847126}
T.~Cazenave and L.~V{\'a}zquez, {\em Existence of localized solutions for a
  classical nonlinear {D}irac field\/}, Comm. Math. Phys. {\bf 105} (1986), pp.
  35--47.

\bibitem[EGS96]{MR1386737}
M.~J. Esteban, V.~Georgiev, and E.~S{\'e}r{\'e}, {\em Stationary solutions of
  the {M}axwell-{D}irac and the {K}lein-{G}ordon-{D}irac equations\/}, Calc.
  Var. Partial Differential Equations {\bf 4} (1996), pp. 265--281.

\bibitem[GN74]{PhysRevD.10.3235}
D.~J. Gross and A.~Neveu, {\em Dynamical symmetry breaking in asymptotically
  free field theories\/}, Phys. Rev. D {\bf 10} (1974), pp. 3235--3253.

\bibitem[GNS95]{MR1331696}
Y.~Guo, K.~Nakamitsu, and W.~Strauss, {\em Global finite-energy solutions of
  the {M}axwell-{S}chr\"odinger system\/}, Comm. Math. Phys. {\bf 170} (1995),
  pp. 181--196.

\bibitem[GSS87]{MR901236}
M.~Grillakis, J.~Shatah, and W.~Strauss, {\em Stability theory of solitary
  waves in the presence of symmetry. {I}\/}, J. Funct. Anal. {\bf 74} (1987),
  pp. 160--197.

\bibitem[KK07]{ubk-arma}
A.~I. Komech and A.~A. Komech, {\em Global attractor for a nonlinear oscillator
  coupled to the {K}lein-{G}ordon field\/}, Arch. Ration. Mech. Anal. {\bf 185}
  (2007), pp. 105--142.

\bibitem[KK08]{ukr-hp}
A.~I. Komech and A.~A. Komech, {\em Global attraction to solitary waves for
  {K}lein-{G}ordon equation with mean field interaction\/}, Ann. Inst. H.
  Poincar\'e Anal. Non Lin\'eaire {\bf 26} (2008), pp. 855--868.

\bibitem[KK09]{ukk-jmpa}
A.~I. Komech and A.~A. Komech, {\em Global attractor for the {K}lein-{G}ordon
  field coupled to several nonlinear oscillators\/}, J. Math. Pures Appl.
  (2009), to appear.

\bibitem[Kom95]{MR1359949}
A.~I. Komech, {\em On stabilization of string-nonlinear oscillator
  interaction\/}, J. Math. Anal. Appl. {\bf 196} (1995), pp. 384--409.

\bibitem[Kom99]{MR1726676}
A.~I. Komech, {\em On transitions to stationary states in one-dimensional
  nonlinear wave equations\/}, Arch. Ration. Mech. Anal. {\bf 149} (1999), pp.
  213--228.

\bibitem[KS00]{MR1748357}
A.~I. Komech and H.~Spohn, {\em Long-time asymptotics for the coupled
  {M}axwell-{L}orentz equations\/}, Comm. Partial Differential Equations {\bf
  25} (2000), pp. 559--584.

\bibitem[KSK97]{MR1434147}
A.~I. Komech, H.~Spohn, and M.~Kunze, {\em Long-time asymptotics for a
  classical particle interacting with a scalar wave field\/}, Comm. Partial
  Differential Equations {\bf 22} (1997), pp. 307--335.

\bibitem[LG75]{PhysRevD.12.3880}
S.~Y. Lee and A.~Gavrielides, {\em Quantization of the localized solutions in
  two-dimensional field theories of massive fermions\/}, Phys. Rev. D {\bf 12}
  (1975), pp. 3880--3886.

\bibitem[MS72]{MR0303097}
C.~S. Morawetz and W.~A. Strauss, {\em Decay and scattering of solutions of a
  nonlinear relativistic wave equation\/}, Comm. Pure Appl. Math. {\bf 25}
  (1972), pp. 1--31.

\bibitem[Sch26a]{sch79361}
E.~Schr{\"o}dinger, {\em Quantisierung als eigenwertproblem\/}, Ann. d. Phys.
  {\bf 79} (1926), pp. 361--376.

\bibitem[Sch26b]{Sch81109}
E.~Schr{\"o}dinger, {\em Quantisierung als eigenwertproblem\/}, Ann. d. Phys.
  {\bf 81} (1926), p. 109.

\bibitem[Sch51]{PhysRev.84.1}
L.~I. Schiff, {\em Nonlinear meson theory of nuclear forces. {I}. {N}eutral
  scalar mesons with point-contact repulsion\/}, Phys. Rev. {\bf 84} (1951),
  pp. 1--9.

\bibitem[Seg63]{MR0153967}
I.~E. Segal, {\em The global {C}auchy problem for a relativistic scalar field
  with power interaction\/}, Bull. Soc. Math. France {\bf 91} (1963), pp.
  129--135.

\bibitem[Str68]{MR0233062}
W.~A. Strauss, {\em Decay and asymptotics for $\square u = f(u)$\/}, J.
  Functional Analysis {\bf 2} (1968), pp. 409--457.

\bibitem[SV86]{MR848095}
W.~A. Strauss and L.~V{\'a}zquez, {\em Stability under dilations of nonlinear
  spinor fields\/}, Phys. Rev. D (3) {\bf 34} (1986), pp. 641--643.

\bibitem[Tit26]{titchmarsh}
E.~Titchmarsh, {\em The zeros of certain integral functions\/}, Proc. of the
  London Math. Soc. {\bf 25} (1926), pp. 283--302.

\bibitem[Yos80]{MR617913}
K.~Yosida, {\em Functional analysis\/}, vol. 123 of {\em Grundlehren der
  Mathematischen Wissenschaften [Fundamental Principles of Mathematical
  Sciences]\/}, Springer-Verlag, Berlin, 1980, sixth edn.

\end{thebibliography}

\def\cprime{$'$} \def\cprime{$'$} \def\cprime{$'$} \def\cprime{$'$}
  \def\cprime{$'$} \def\cprime{$'$} \def\cprime{$'$} \def\cprime{$'$}
  \def\cprime{$'$} \def\cprime{$'$}

\end{document}